\documentclass[sigconf]{acmart}
\AtBeginDocument{%
  }

\copyrightyear{2025}
\acmYear{2025}
\setcopyright{cc}
\setcctype{by}
\acmConference[DIS '25]{Designing Interactive Systems Conference}{July 5--9, 2025}{Funchal, Portugal}
\acmBooktitle{Designing Interactive Systems Conference (DIS '25), July 5--9, 2025, Funchal, Portugal}
\acmDOI{10.1145/3715336.3735817}
\acmISBN{979-8-4007-1485-6/2025/07}
\begin{document}

\title[Insights from Designing Context-Aware Meal Preparation Assistance for Older Adults with MCI and Their Care Partners]{Insights from Designing Context-Aware Meal Preparation Assistance for Older Adults with Mild Cognitive Impairment (MCI) and Their Care Partners}

% \affiliation{%

\author{Szeyi Chan}
\authornote{Both authors contributed equally to this research.}
\email{chan.szey@northeastern.edu}
\affiliation{%
  \institution{Northeastern University}
  \country{USA}
}

\author{Jiachen Li}
\authornotemark[1]
\email{li.jiachen4@northeastern.edu}
\affiliation{%
  \institution{Northeastern University}
  \country{USA}
}

\author{Siman Ao}
\email{sao8@gatech.edu}
\affiliation{%
  \institution{Georgia institute of technology}
  \country{USA}
}

\author{Yufei Wang}
\email{ywang3283@gatech.edu}
\affiliation{%
  \institution{Georgia Institute of Technology}
  \country{USA}
}

\author{Ibrahim Bilau}
\email{ibilau3@gatech.edu}
\affiliation{%
  \institution{Georgia Institute of Technology}
  \country{USA}
}

\author{Brian Jones}
\email{brian.jones@imtc.gatech.edu}
\affiliation{%
  \institution{Georgia Institute of Technology}
  \country{USA}
}

\author{Eunhwa Yang}
\email{eunhwa.yang@design.gatech.edu}
\affiliation{%
  \institution{Georgia Institute of Technology}
  \country{USA}
}

\author{Elizabeth D Mynatt}
\email{e.mynatt@northeastern.edu}
\affiliation{%
  \institution{Northeastern University}
  \country{USA}
}

\author{Xiang Zhi Tan}
\email{zhi.tan@northeastern.edu}
\affiliation{%
  \institution{Northeastern University}
  \country{USA}
}

\renewcommand{\shortauthors}{Chan and Li, et al.}

\begin{abstract}
    Older adults with mild cognitive impairment (MCI) often face challenges during meal preparation, such as forgetting ingredients, skipping steps, or leaving appliances on, which can compromise their safety and independence. Our study explores the design of context-aware assistive technologies for meal preparation using a user-centered iterative design process. Through three iterative phases of design and feedback, evolving from low-tech lightbox to a digital screen, we gained insights into managing diverse contexts and personalizing assistance through collaboration with older adults with MCI and their care partners. We concluded our findings in three key contexts--routine-based, real-time, and situational--that informed strategies for designing context-aware meal prep assistance tailored to users' needs. Our results provide actionable insights for creating technologies to assist meal preparation that are personalized for the unique lifestyles of older adults with MCI, situated in the complex and dynamic homebound context, and respecting the collaboration between older adults and their care partners.
\end{abstract}



\begin{teaserfigure}
  \includegraphics[width=\textwidth]{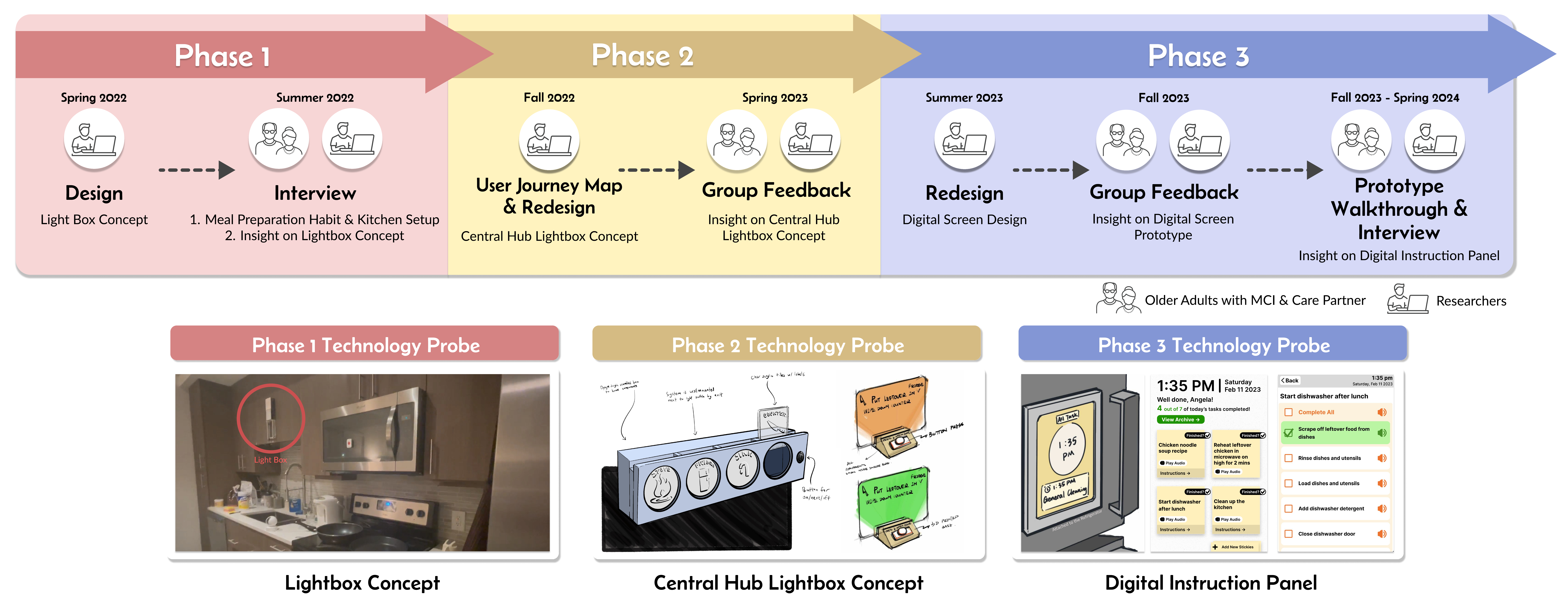}
  \caption{An overview of our three-phase iterative study conducted from Spring 2022 to Spring 2024. Each phase included a technology probe tailored to support meal preparation for older adults with MCI and their care partners. Phase 1 introduced the Lightbox Concept, Phase 2 introduced the Central Hub Lightbox Concept, and Phase 3 introduced the Digital Instruction Panel. This shows the study timeline and the progression of design iterations throughout the study.}
  \label{fig:teaser}
  \Description{The figure provides a visual overview of a three-phase iterative study conducted from Spring 2022 to Spring 2024. Each phase included a technology probe tailored to support meal preparation for older adults with MCI and their care partners.  The top section shows a timeline divided into three phases: Phase 1 (Spring 2022 to Summer 2022) introduced the Lightbox Concept, Phase 2 (Fall 2022 to Spring 2023) introduced the Central Hub Lightbox Concept, and Phase 3 (Summer 2023 to Spring 2024) introduced the Digital Instruction Panel. Each phase is represented with corresponding activities, including design, user feedback, and prototype walkthroughs. The bottom section visually depicts the technology probes developed for each phase: Phase 1 shows the Lightbox mounted near a microwave; Phase 2 shows a sketched prototype of the Central Hub Lightbox with multiple icons representing kitchen tasks; Phase 3 displays a digital screen prototype with task instructions and completion tracking.}
\end{teaserfigure}


\maketitle

\section{Introduction}
Older adults with mild cognitive impairment (MCI) experience cognitive decline more progressively than expected for their age group~\cite{gauthier2006mild, petersen2016mild}. Approximately 10–20\% of individuals over the age of 65 are affected by MCI~\cite{mci_prevalence}. While older adults with MCI are able to live independently and perform various daily activities, they sometimes encounter challenges with more complex tasks such as forgetting appointments, leaving the stove on, or struggling to follow a recipe during meal preparation.  These challenges increase the risk of accidents and reduce their quality of life.
Care partners, often family members or friends, play an important role in supporting older adults with MCI to live at home, but this responsibility can lead to stress and burnout due to the workload. %~\cite{}. 
Assistive technologies have the potential to reduce this burden by offering direct support to older adults with MCI while enabling them to maintain a greater sense of autonomy~\cite{mccarron2019web, dada2024intelligent,lobe2022empowering}. For example, reminder applications assist with daily routines like medication management~\cite{mathur2022collaborative}, food intake~\cite{o2011video}, and cooking~\cite{hackett2022remind}, significantly helping older adults with MCI maintain daily activities.
For these assistive technologies, one of the key factors for successful integration into seniors' lives is context awareness~\cite{holthe2022digital, maier2015remember,hackett2022remind}.

Context-aware systems, which respond to users' surroundings, can enhance assistive technologies by providing situation-specific support~\cite{satyanarayanan2001pervasive} has the potential to assist older adults with MCI and their care partners. These context-aware assistive technologies could serve as ``compensatory support'' that empowers them to maintain functionality at the level they had before the decline of cognitive abilities~\cite{zubatiy2021empowering,zubatiy2023distributed}. Despite their potential, designing context-aware assistive systems remains a significant challenge, in part due to the need for personalization in understanding the variability of environments and the diverse needs of users, such as older adults with MCI who may require different levels of support~\cite{mathur2022collaborative, klakegg2021care}. Furthermore, any support requires further personalization by the care partner as the situation and needs of the adults with MCI differ between households. 

Among all the contexts, meal preparation is a typical yet complicated and dynamic daily routine for older adults with MCI.
It is one of the five instrumental activities of daily living~\cite{lawton1969iadl} and an indicator of one's ability to age at home. 
Yet older adults with MCI often face challenges in meal preparation due to memory decline, making it difficult to manage multiple steps, ensure safety, and follow routines. While assistive technologies have the potential to support them~\cite{zubatiy2021empowering}, the effectiveness of these systems depends on their ability to identify and manage the various contexts relevant to this task to provide just-in-time support. 
Meal preparation involves managing multiple tasks, such as ingredient preparation and following cooking instructions~\cite{craik2006planning}. While previous research has explored technological support through interactive guidance and smart kitchen appliances~\cite{chan2023mango,blasco2014smart,hamada2005cooking,neumann2017kognichef}, little is known about the various contexts important for meal preparation with older adults with MCI. 
Integrating context awareness with personalization adds further complexity, as the symptoms of MCI vary and progress over time, altering individual habits and needs. 

Moreover, the collaborative nature of meal preparation between people with MCI and their care partners complicates the design of assistive systems. 
Meal preparation tasks are inherently collaborative and often involve coordination between family members in the same household. 
With the decline in cognitive ability due to MCI, the task distribution related to cooking might change significantly. 
Care partners may become more involved in the cooking task and take on more responsibility due to multiple factors such as safety concerns~\cite{yu2023dyadic}. 
This presents a unique issue for the population of older adults with MCI who are aging in place, as they still maintain some functionality compared to individuals in the late stages of dementia who lost the ability to cook independently~\cite{papachristou2013impact}. 
Despite the importance of this issue, limited studies incorporate care partners in the design process for meal preparation activities considering the specific challenges and needs of older adults with MCI.
We believe that the changes in functionality alter the original collaboration process in meal preparation tasks that technology designed to assist this process must take it into consideration. 

Our research addresses these gaps by investigating the meal preparation needs for older adults with MCI and their care partners and how a context-aware device could assist in meal preparation. We employed a participatory and iterative design methodology collaboratively with both older adults with MCI and their care partners, ensuring that the system remains grounded in their lived experiences. Care partners play an important role in shaping the design of assistive technologies for older adults with MCI. Through their caregiving experience, they provide valuable insights into the contexts that assistive systems should address. Collaborative design with care partners ensures that these technologies prioritize safety and independence for older adults while also potentially supporting the caregiving dynamic. Moreover, designing devices to be accessible to the entire household fosters inclusivity and usability, enabling shared spaces and collaborative engagement. 

Over three phases of design and feedback, we iteratively refined our technology probe, progressing from a low-tech lightbox device to a context-aware personalized solution. From this iterative process, we summarize the takeaway as three critical contexts for assistive systems: routine-based, real-time, and situational. These insights informed the development of adaptive assistance strategies tailored to users' needs. By focusing on supporting older adults with MCI in meal preparation scenarios, this study addresses challenges by identifying essential contexts and offering actionable insights for designing assistive technologies that meet the needs of both older adults with MCI and their care partners.

The primary contributions of this work are threefold. First, we address the research questions: \textbf{(Q1) What context should an application be aware of when supporting MCI older adults during meal preparation?} and \textbf{(Q2) How can a personalized meal preparation reminder device be designed to function effectively within these contexts?} Second, by iterating from a low-tech device to a context-aware personalized system, we provide actionable insights for designing assistive technologies that meet the unique challenges faced by older adults with MCI. These findings lay the foundation for future development of context-aware personalized systems, supporting not only meal preparation but also broader daily activities to enhance independence and safety for this population. Finally, we reflect on the design process, highlighting key lessons learned for creating effective meal preparation assistive technologies for older adults with MCI.

\section{Related Work}
\subsection{Meal Preparation Assistant}

Meal preparation is an everyday task that involves multiple steps, such as preparing raw ingredients, managing time, multitasking, and considering factors like ingredient availability and nutrition. These tasks collectively demand significant cognitive effort. To reduce the challenges associated with these steps, researchers and companies have explored various technological solutions to simplify the meal preparation process.

Cooking guidance has been a focus, with tools designed to provide step-by-step instructions and real-time support. Existing commercialized voice assistants, such as Amazon Alexa and Google Home, are able to leverage verbal communication to provide basic support such as recipe searches and setting reminders. Providing proactive real-time support remains an active area of research~\cite{chan2023mango, sciuto2018hey, frummet2024cooking}.
For example,~\citet{chan2023mango} recently explored the use of large language models with voice assistants to deliver detailed instructions and answer users' wide range of queries during cooking. Furthermore, systems like ``Cooking Navi'' integrate multimodality, such as text, video, and audio, to assist users in managing recipes and tasks~\cite{hamada2005cooking}. While these tools help users navigate complex recipes effectively, they often struggle to adapt to the surrounding context in the kitchen and lack the capability to address real-time physical challenges that may arise during cooking.

To complement these limitations, researchers have also examined the potential of smart kitchen appliances to enhance meal preparation. Smart refrigerators, for instance, can monitor stored items and provide real-time inventory updates to users' phones~\cite{miniaoui2019introducing, morris2021inventory,ayub2022don}. 
Other research has explored the integration of Internet of Things technologies into kitchen systems to provide tailored support during cooking, such as helping users monitor and adjust cooking tasks dynamically~\cite{angara2017foodie, mennicken2010first, sugiura2010cooking, weber2023designing, chen2010smart}. 

Prior work has also explored how to provide support for users with vision loss, memory difficulties, trouble following complex instructions, or physical limitations such as reduced mobility or strength~\cite{li2024recipe, li2021non, kuoppamaki2021designing, bouchard2020smart, kosch2019digital}. For example, ~\citet{kosch2019digital}'s ``Digital Cooking Coach'' integrated visual and auditory step-by-step instructions, allowing users to stay on task and complete meals accurately. Similarly, ~\citet{li2024recipe} explored design features of assistive technologies that could enhance recipe access for individuals with visual impairments, such as voice-guided instructions, tactile feedback, and simplified recipe formats.

Although these tools offer valuable features, they are typically designed to address a single aspect of the meal preparation process. For instance, one may guide users through recipes but lack support for planning and ingredient management, cooking and safety monitoring, and post-meal preparation clean up. 
In this work, we aim to address these limitations by designing a context-aware meal preparation assistant that can recognize the ongoing context of kitchen activities and provide timely, relevant assistance tailored to the user's needs, specifically for older adults with MCI.

\subsection{Assistive Tools and Technologies for Older Adults with MCI}
Older adults with MCI face unique challenges in managing daily tasks due to memory decline and reduced cognitive abilities~\cite{gauthier2006mild,petersen2016mild}. While aging in place offers numerous benefits, such as reduced reliance on care facilities and greater autonomy, it also presents significant challenges for maintaining daily routines for tasks like meal preparation, medication reminders, and household chores~\cite{mynatt2000increasing}.

To manage these challenges, many older adults with MCI rely on their caregivers for reminders and assistance in daily activities~\cite{paradise2015caregiver, ryan2010caregiver,beatie2021caregiver}. In addition, “low-tech” tools like sticky notes, calendars, and physical reminders are widely used~\cite{mathur2022collaborative,raghunath2020creating}. These tools are easy to use, require no technological setup, and are highly accessible. However, physical reminders can be easily misplaced and forgotten, and as memory declines, locating and interpreting these reminders can become increasingly difficult~\cite{raghunath2020creating}. Despite these limitations, such tools remain a common strategy for maintaining routines and cognitive support.

Technological advancements have introduced a range of assistive tools for older adults. \citet{pollack2005intelligent} outlined the three primary goals of assistive technology for older adults with MCI are to ensure their safety while notifying caregivers when risks arise, to assist them in performing daily tasks, and to assess their cognitive status. General-purpose tools like voice assistants and reminder applications, including Amazon Alexa, Google Home, and mobile apps, partially address these goals by offering verbal prompts, alarms, and grocery list management\cite{bimpas2024leveraging}. These features make them popular among older adults with MCI, as they help with memory-related tasks and reduce cognitive burdens.

In addition to the general-purpose technologies, researchers have developed specialized tools for older adults, such as calendaring systems for managing schedules~\cite{zubatiy2023distributed}, conversational AI systems~\cite{mathur2023did}, and health check-in applications~\cite{mathur2022collaborative}. These tools help reduce cognitive load by prompting reminders and step-by-step instructions, enabling users to perform daily tasks more independently. 

Meal preparation poses unique challenges for older adults with MCI because it involves managing multiple interconnected tasks, such as tracking cooking times, remembering ingredients, and safely using appliances~\cite{yaddaden2020using, johansson2011cognitive}. Assistive technology can be particularly beneficial in this context by helping users manage these tasks more effectively. Researchers have explored innovative designs to support cognitively impaired users in cooking~\cite{johansson2011cognitive, jonsson2019reminder}. For example, ~\citet{bouchard2014smart} developed a smart range prototype that uses load cells, heat sensors, and electromagnetic contacts embedded in the range to monitor and guide users during meal preparation.

Despite these advancements, many current tools fail to fully achieve the primary goals of assistive technology, particularly in ensuring safety, assisting with task completion, and notifying caregivers when necessary~\cite{pollack2005intelligent}. In this work, we aim to develop a solution that aligns with these goals by understanding the contextual needs of older adults with MCI and providing reminders, tailored guidance, and dynamic support specifically during meal preparation.

\subsection{Context Awareness in Assistive Technology}
~\citet{abowd1999towards} defines context as “Context is any information that can be used to characterize the situation of an entity. An entity is a person, place, or object that is considered relevant to the interaction between a user and an application, including the user and applications themselves.” Context awareness enables assistive technologies to recognize a user's state and environment and adapt their behavior to provide relevant support~\cite{satyanarayanan2001pervasive}. Sensors are often utilized for recognizing and interpreting contexts in these systems~\cite{shrinidhi2022novel, kosch2018smart}.

Context awareness has been widely explored in assistive technologies, particularly in healthcare~\cite{fallahzadeh2016context, anzanpour2016context, kulkarni2011mphasis, kyriazakos2016ewall, chang2016context}. For example, wearable devices such as health watches leverage bio-signals like heart rate, blood pressure, respiration, and body temperature to develop remote health monitoring services~\cite{kang2006wearable}. Similarly, smart home technologies incorporate sensors and IoT devices to monitor users' activities and ensure safety and comfort in daily living~\cite{shrinidhi2022novel, stanford2002using, mynatt2000increasing}. Early research also demonstrated the value of context awareness in assistive systems for older adults. Autominder~\cite{pollack2003autominder} provided adaptive reminders for daily activities by using contextual information like user's daily plan, event times, and durations. Similarly, the Independent Lifestyle Assistant~\cite{haigh2006independent} focused on monitoring activities like medication and mobility, combining sensor data with machine learning to issue alerts and provide assistance to caregivers. 

While these studies demonstrate the effectiveness of context-aware systems in general assistive applications, there are unique challenges in meal preparation, particularly for older adults with MCI. These challenges include difficulties in remembering steps, managing ingredients, and using appliances safely. As a result, the kitchen context becomes more complicated, involving tracking cooking times, monitoring ingredient usage, managing appliance status, and ensuring safety during multi-step processes. Existing systems often address a single aspect of meal preparation, such as using a smart range to monitor and guide cooking steps~\cite{bouchard2014smart}. However, these tools lack a comprehensive understanding of all the contexts associated with meal preparation.

In our study, we aim to identify and analyze the various contexts that occur in the kitchen during meal preparation by gathering feedback through our meal preparation assistive technologies design prototypes. By understanding these contexts, our result provide a foundation for designing context-aware assistive technologies that provide tailored functionality, helping older adults with MCI reduce the challenges associated with meal preparation.

\section{Research Approach Overview }
We took an iterative design approach to understand the problem and design our meal preparation assistant. This resulted in three iterative phases, with each phase iteratively refined based on the feedback gathered from the previous phase. Through the sessions with older adults with MCI, we identified evolving preferences and needs that guided each iteration. Figure~\ref{fig:teaser} shows the overview of our research approach and timeline. %This iterative design methodology allowed us to incorporate user feedback to refine our designs.
\begin{description}
    \item[Phase 1: Technology Probe and Interviews]\leavevmode\\
    The first phase began by exploring participants' meal preparation habits and kitchen setup and organization to gain insight into their routines and environments. Additionally, we introduced our lightbox concept through a pre-recorded usage video as a technology probe. We then interviewed participants to gather feedback on the lightbox's functionality and identify areas for improvement.
    
    \item[Phase 2: Iterative Prototyping and Group Feedback]\leavevmode\\
    The second phase involved refining our lightbox concept into its second iteration as a central hub. We then conducted a group feedback session with participants to gather insights for the second prototype.
    
    \item[Phase 3: Refining and Testing with a Technology Probe]
    The third phase involved creating a digital instruction panel, which was built upon the findings from the second phase. We first ran another group feedback session to refine the design. We then conducted a prototype walkthrough, during which participants interacted with the digital panel and provided feedback through interviews to evaluate the prototype's effectiveness and collect insights for this iteration.
\end{description}
In the following sections, we will describe each phase in depth and how user feedback influences the evolution of the prototype.

% --- In different file ----- %
\section{Phase 1: Technology Probe and Interviews}
Our first phase involved an in-depth interview about participants' meal preparation habits and kitchen setup. We also gathered participants' reactions and feedback to smart reminder systems for meal preparation using a simple low-tech technology probe -- ``lightbox''.

\begin{figure*}
  \includegraphics[width=\textwidth]{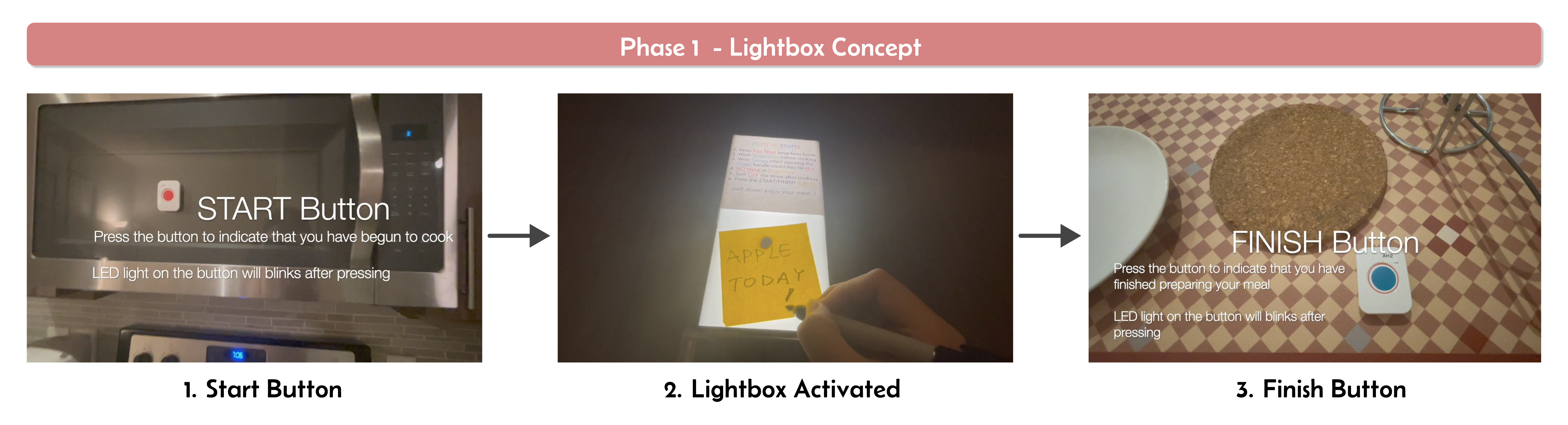}
  \caption{The interaction flow for the Lightbox concept. The process begins by pressing the Start button to initiate the system. The Lightbox then illuminates, displaying customized instructions or reminders for the user. Once meal preparation is complete, the user presses the Finish button, located on the dining table, to indicate task completion.}
  \label{fig:phase1}
  \Description{The figure illustrates the interaction flow for the Lightbox concept in three steps. Step 1 shows the Start button, located on a microwave, which users press to begin the process. Step 2 shows the Lightbox illuminating, displaying customized instructions or reminders written on a sticky note attached to the light. Step 3 shows the Finish button, located on the dining table, which users press to indicate that meal preparation is complete. Both buttons have an LED light that blinks after being pressed, providing visual feedback.}
\end{figure*}

\subsection{Lightbox Technical Probe Design}
We prototyped a simple context-aware reminder device -- lightbox -- that guides users to pre-selected instructions when they decide to start a meal preparation activity (shown in Figure~\ref{fig:phase1}).
Its low-tech design was motivated by prior research showing that older adults with MCI often struggle with complex, multi-step digital interfaces~\cite{schmidt2019predictors, chen2022interface}. We incorporated familiar physical buttons to control the device to reduce the learning curve. Pressing the start button activates the lightbox, which lights up immediately to provide attention-grabbing visual feedback without requiring digital literacy. This interaction minimizes cognitive load and aligns with users’ existing routines.
The prototype allows us to probe the user about what information they want to receive and how they will interact with reminder technology in the kitchen. The lightbox design consisted of three key components: a start button, an end button, and a rectangular lightbox with a surface for placing pre-selected written reminders. We intentionally avoided advanced technology to ensure simplicity for older adults with MCI. 
The lightbox was portable, allowing participants to position it in locations that were most convenient and accessible for them. 

To begin a task like cooking, participants could first press the start button,then an LED light on the button would blink, activating the lightbox and signaling that the task had begun. During the task, the surface of the lightbox would display customized cooking instructions or reminders that participants and carepartner had prepared in advance. Once the task was completed, participants could press the end button to indicate they were finished. An LED light on the end button would blink to confirm that it had been pressed, and the lightbox would turn off. 
The design of the lightbox mimics an interaction of user-initiated check-ins, which has been proven successful with older adults with MCI in managing routines~\cite{mathur2022collaborative}.

\subsection{Participants and Procedures}
In this phase of the study, we conducted interviews with fourteen participants recruited through a senior lifestyle program for adults with MCI and their care partners, Cognitive Empowerment Program (CEP). 
Participants were contacted via text messages and emails to inquire about their interest in participating.
This study involved seven pairs of participants, each consisting of an older adult with MCI and their care partner.% 
Each interview session lasted approximately 60 minutes and was conducted via Zoom while the participants were in their own residences. Screenshots of participants’ kitchen spaces were taken during the interview sessions, and the interview was recorded for further analysis. Prior to participation, both the adult with MCI and their care partner signed consent forms that outlined the study’s purpose and obtained their permission to record and take screenshots. This study was approved by the university's institutional review board. 

The interviews were designed to explore several topics, including the participants’ backgrounds in meal preparation, the setup and organization of their kitchens, and their reactions to the Lightbox technology probe which was demonstrated through a pre-recorded video. Additionally, participants were guided through meal preparation scenarios in their own kitchen to discuss how the system could integrate into their daily routines and provide support.
 
Three researchers organized the data collected from these interviews to capture key observations. Then, they categorized and analyzed it using an affinity diagram created in Miro to identify common themes and patterns across participants’ responses. This phase provided insights into participants’ meal preparation habits and initial thoughts on a meal preparation reminder device.

\subsection{Results}
\subsubsection{Insight from the Cooking Habit Interview}
Through the interviews, we observed that while older adults with MCI have diverse cooking habits, they share certain commonalities in their needs and routines. We categorized the shared needs into three key contexts: \textbf{routine-based, real-time, and situational contexts}. This categorization aligns with our overarching goal of understanding the contextual factors critical for designing context-aware personalized assistance systems. Furthermore, the interviews provided answers to key questions for designing meal prep reminder assistance that addresses the unique challenges of supporting older adults with MCI during meal preparation.

\paragraph{\textbf{Routine-Based Contexts}}
Recurring habits, preferences, or practices were often highlighted by older adults with MCI as an area which care partners often have to provide reminders for. We define triggers for reminders based on recurring practices as \textbf{Routine-Based Context}. 
For older adults with MCI, routines around meal preparation involve not just cooking itself, but many recurring events closely associated with cooking, including health- and diet-related tasks as well as grocery-related activities, where reminders are often needed.
Many of these reminders come from or are set by the care partners.
For instance, some care partners have adopted technology to assist with health-related routines associated with food intake: \textit{``He used to have trouble remembering to take blood pressure, so we set a reminder on Google Home, which helped condition himself to establish the habit.’’} (by partner of P1-3). Dietary and nutritional needs were also highlighted as another significant routine-based reminder. P1-3 who avoid salt shared: \textit{``I'm watching the sodium content from the meal kit to make sure they have maybe around 500 mg per serving, not too high.’’}. Reminders need to understand the meal preparation and daily snacking routines of the user and provide timely reminders.

Grocery management was another important routine discussed by participants, with some older adults with MCI using assistive technologies to help manage this task independently. P1-1 described their process: \textit{``We use Alexa a lot. I was sitting at the Alexa to create the shopping list. To go through our whole routine, he stands at the pantry and refrigerator and looks in there. And then say what we need, so I'll put all those on the list. Then I'll have Alexa print this shopping list.}’’ Similarly, P1-2 shared: \textit{``Google Home does keep the grocery list which I appreciate a lot. When I need something I can just tell her and then when I go to the grocery store, there it is.’’}. Existing smart assistants allowed older adults with MCI to handle grocery management independently.

\paragraph{\textbf{Real-Time Contexts}}
Meal preparation often involves actions that need to be taken at critical moments. We define triggers for reminders that depend on these moments as \textbf{Real-Time Contexts}.

Participants described challenges in real-time decision-making and error prevention during meal preparation. P1-2 shared, \textit{``With MCI, I've started cooking but I forget to put ingredients that I should have put into.’’} Mistakes in cooking sequences were a common problem and led to safety concerns. For example, P1-4 described a time when using a microwave for cooking: \textit{``The only problem he had is he cooked rice one time and cooked it way too long and burned the rice, and it got so hot. It actually cracked the bowl. I think the problem was that I'm supposed to cook on high for five minutes. And then I'm supposed to cook it on low for 20 minutes. And I forgot to press the low.’’} Real-time reminders based on the current cooking step could potentially avoid mistakes that could lead to safety concerns. These reminders should also be constructed from the dyads' experiences based on the person with MCI's past mistakes.

In addition to error prevention, participants discussed the need for real-time task coordination, such as when different parts of the meal should be prepared. P1-2 explained, \textit{``I need to think the order of cooking to decide how to get it ready at the same time. If I make my own sauce, I need to start it before cooking noodles; if I use bottled sauce, I need to remember getting the stool to get stuff out from top shelf.’’} Managing task sequences is a cognitively demanding aspect of meal preparation, particularly when multitasking, and it becomes even more challenging for older adults with MCI.

Participants also shared their current workarounds for managing real-time cooking tasks, such as relying on verbal instructions from existing assistive technologies or setting multiple reminders across different devices. For instance, P1-3 shared \textit{``We have Google Home in the kitchen that I often use as timers. I will use microwave timer, but when it's already in use and I need more timers, I will use Google Home. There's no limit on how many timers you can run.’’} When recipes or specific instructions were forgotten, participants turned to external resources, such as recipe boxes or written boards. P1-7 shared, \textit{``My board is my list telling me how long to cook things, what temperature to set.’’} For more detailed guidance, they accessed additional resources. P1-7 added, \textit{``Sometimes I forget how to make something, I go to my recipe box and pull out the instructions.’’} Similarly, P1-3 explained, \textit{``I sometimes will ask questions about an ingredient or a nutritional fact with Google Home.’’}

\paragraph{\textbf{Situational Contexts}}
In addition to task-specific contexts, participants also expressed the need for reminders to react to the physical situation in the kitchen. We called these triggers \textbf{Situational Context}. Participants shared various scenarios related to their surroundings, including safety concerns, appliance use, and environmental awareness. First of all, with declining cognitive abilities, older adults with MCI require increased support to ensure kitchen safety during meal preparation, particularly when operating appliances. Almost all the participants have experienced some close calls with the kitchen appliances. P1-4 shared: \textit{``Sometimes we'll leave a burner on, especially because when we turn it way down, it's hard to tell that it's even on. A couple years ago member was cooking on stove and it burned and set off the smoke alarm. But we haven't had the problem for years.''}

Some participants opted for simpler appliances, such as microwaves, to reduce the fire risk and relied on assistive technologies like setting reminders to help prevent incidents. However, challenges persist. For example, P1-7 explained, \textit{``My biggest issue is staying in the kitchen while I'm cooking. I can set up the reminder but if I'm in the other part of my apartment, sometimes I couldn't hear it and I may burn something.’’} These experiences highlight the difficulties older adults with MCI face during meal preparation, even when they attempt to use workarounds. 

Beyond safety, participants expressed the need for assistance in understanding and operating kitchen appliances. P1-6 explained, \textit{``He doesn’t know when to use toast or other functions, and when he failed, he burns things’’} Likewise, P1-1 shared, \textit{``The other thing I use the internet for is if I need a manual since I don't remember exactly, so I'll go get the manual.’’}

Situational contexts also include environmental factors, such as ensuring the kitchen is properly cleaned after use. Currently, care partners often handle these tasks when they notice something has been missed. For instance, P1-4 described, \textit{``She (care partner) reminds stuff that I forgot to do, making sure the floor and tabletop is clean, etc. Sometimes you have to put your hand on the table to make sure it's clean since eyes cannot see it.’’} In addition, P1-3 mentioned minor issues like forgetting to close cabinet doors: \textit{``This is minor but sometimes he forgot to close the cabinet doors. I just come along behind and close it. If there's something beeps when we left the drawers or cabinet doors open that might be helpful.’’} While these tasks may not pose immediate safety concerns, assistive systems providing environmental reminders, such as alerts for unclosed drawers or unclean surfaces, could promote independence in organizing the kitchen environment.

Food storage and freshness were also important situational contexts highlighted by participants. Cognitive decline can make it difficult to remember how long food has been stored or to recognize signs of spoilage. Participants explained their different approaches to food safety from ``\textit{put date on my leftovers}'' (P1-7) to having ``two day limit for leftovers'' (P1-3).

Finally, situational contexts also involve identifying who is responsible for specific kitchen tasks. In some cases, the care partner handles the cooking and may not require reminders. As P1-1 shared, \textit{``My husband is the cook. I am not. He is preparing meals for some years now. So he does all of that.’’} In contrast, P1-2 stated, \textit{``I prepare dinner every day. We rarely eat out.’’} Therefore, reminder systems can potentially tailored to the appropriate individual, as the needs of care partners often differ from those of older adults with MCI.

\subsubsection{Insight for the Lightbox Technical Probe}
After watching the video, participants found the lightbox could support \textit{\textbf{routine-based contexts}} by providing reminders based on their established habits and preferences. The personalization instructions by the user and care partner helped make the system more practical and engaging for users. For example, some participants appreciated how the system simplified routine actions and reduced cognitive load, while others highlighted its potential to assist with frequently used appliances. P1-6 emphasized the usefulness of the lightbox for routine appliance usage, stating, \textit{``The lightbox might be helpful to tell him how to use a toaster oven, which is a very complex equipment he actually uses often and would mix up.’’} Similarly, P1-7 noted also sees the value of integrating simple, habitual actions of pressing a button into the user’s regular routine.

\textbf{Functionality Desire:} Participants reflected on the lack of support for real-time and situation context as these contexts directly influence the success of meal preparation tasks. \textbf{\textit{Real-time contexts}} involve providing guidance at critical moments, such as reminders to check on food or complete specific steps in a sequence. For instance, P1-5 shared the need for dynamic prompts based on real-time contexts like, \textit{``Don't forget the biscuit is ready in 10 mins’’} or \textit{``When I was making salad dressing or something: to be sure and blend it long enough.’’}  To deliver proactive support in real-time contexts, participants suggested using audio input and output for adding and reading instructions or prompts, rather than relying solely on passive message displays like those on the lightbox. This would help ensure that all instructions are followed without relying on visual cues. As P1-4 noted, \textit{``It would be more useful to have something like Siri or Alexa reading the recipe out rather than having something that I would have to read.’’} %Addressing these contexts, which are currently absent in the lightbox design, would help reduce cognitive load and enhance the safety and efficiency of meal preparation for older adults with MCI.

Participants emphasized including more \textbf{\textit{situational contexts}} that near the user’s surroundings can further enhance the functionality of the lightbox concept. Firstly, participants expressed the value of dynamic, "always-on" notes tailored to specific tasks during meal preparation, such as \textit{``Clean up the kitchen’’ and ``Make sure the oven is off, stove is off’’} (P1-5) to help not forget. Secondly, participants noted the potential for assistive technologies to detect and respond to running low on specific items during cooking. P1-4 suggested, \textit{``You could tell it (the lightbox) we are running out of something to help with ordering grocery,’’} while P1-7 added, \textit{``Remember to buy more eggs.’’} These contexts from the surroundings would help streamline grocery management and reduce the cognitive load associated with tracking inventory. However, while the lightbox was designed for older adults with MCI and their care partners, P1-1 raised concerns about its ability to distinguish between users and ensure reminders are directed to the appropriate individual. Lastly, P1-3 mentioned, \textit{``The instruction on the lightbox is hard to read now. Especially for people with limited vision or having trouble looking into the light.’’}, pointing out that the readability of the lightbox was another important factor for them. Additionally, both P1-6 and P1-3 suggested incorporating status lights into the buttons, with green indicating start and red indicating stop, to improve usability. Overall, the lightbox design should account for physical environmental factors, such as lighting conditions, screen design, and text size, to ensure instructions are clear and accessible for all users. 

\textbf{Physical Design Desire:} In addition to its functionality, participants also suggested hardware design improvements. One key suggestion was optimizing the lightbox’s location for better interaction and readability. For example, P1-6 preferred it near the toaster, stating, ``The lightbox ideally should be next or above the toaster so I can see it while I’m using the toaster. I have to watch what's in the toaster anyway, so it's better the lightbox is also near it.’’ 
Another commonly discussed feature was button placement and functionality. Many participants preferred having the buttons integrated directly into the lightbox and having a single button. P1-4 shared, \textit{``I'd prefer to have the button on the lightbox rather than having them separated. If you have a large kitchen it might work, but for our small kitchen, we are trying to minimize the number of things you have out on the surface’’}, where as P1-1 mentioned: \textit{``I would have liked the button you just push it on and you push it off.''}

\begin{figure*}
  \includegraphics[width=\textwidth]{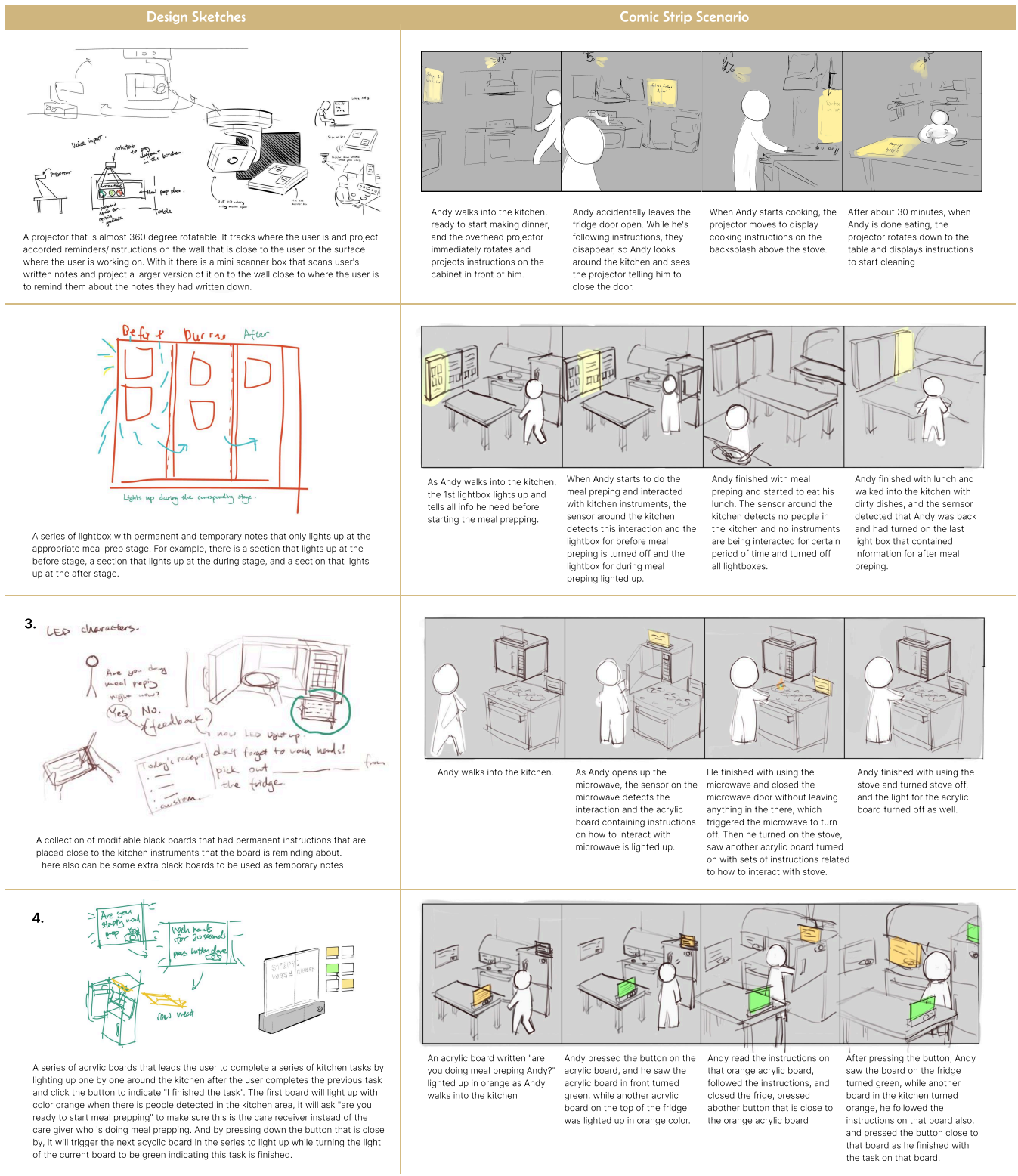}
  \caption{Design sketches and comic strip scenarios illustrating four prototypes for context-aware meal preparation assistance, developed during our brainstorming process for the next prototype iteration. The left column presents conceptual sketches for (1) a 360-degree rotatable projector that displays reminders and instructions, (2) a lightbox system that stages reminders at different meal preparation phases, (3) modifiable blackboards that align with kitchen appliances, and (4) a series of sequential acrylic boards for guiding task completion. The right column shows the corresponding usage scenarios through comic strips, highlighting how each prototype interacts with the user (Andy) during meal preparation tasks.}
  \Description{Design sketches and comic strip scenarios illustrating four prototypes for context-aware meal preparation assistance, developed during our brainstorming process for the next prototype iteration. The left column presents conceptual sketches for (1) a 360-degree rotatable projector that displays reminders and instructions, (2) a lightbox system that stages reminders at different meal preparation phases, (3) modifiable blackboards that align with kitchen appliances, and (4) a series of sequential acrylic boards for guiding task completion. The right column shows the corresponding usage scenarios through comic strips, highlighting how each prototype interacts with the user (Andy) during meal preparation tasks.}
  \label{fig:figure_design_skerches_phase_2}
\end{figure*}

\section{Phase 2: Iterative Prototyping and Group Feedback}
\subsection{Design Process}
After Phase 1, we collaboratively created user journey maps outlining key user actions, design touchpoints, and potential areas for improvement. This mapping process helped us to systematically analyze the feedback gathered during the interviews and identify opportunities for refining the system. From the interview data, we categorized meal preparation instructions into three phases: (1) before meal preparation, (2) during meal preparation, and (3) post meal preparation. To refine our concept, we first brainstormed and proposed four different design directions, each accompanied by a design sketch, a description of the system, technical specifications or limitations, design justification, and a short storyboard capturing key user interactions (shown in Figure~\ref{fig:figure_design_skerches_phase_2}). The researchers then engaged in discussions about the four design sketches and decided to proceed with the final one. We realized that although the first design, a 360--degree rotatable projector, was visually appealing, it was too idealistic for real-world scenarios. The experience relied heavily on the accuracy of the motion detection algorithm and was also constrained by the environment, as not all older adults have spacious kitchens with ample blank space for projecting information. The second design, a light box system, was more plausible but still required a large empty area to place the device, making it more suitable for a large shared kitchen. The last two designs were similar and aligned with the Phase 1 criteria for different cooking tasks. Ultimately, we chose the final design for its extensive customizability. It features a central board presenting all tasks in a sequential order, complemented by detailed instruction boards at key interaction points to guide older adults with MCI step by step. This design incorporated the feedback from Phase 1. We increased the number of tasks and located the instructions closer to the task location. We also simplified the button interaction.

Although our prototype is designed to support all three phases of meal preparation instructions, we chose to focus on the post meal preparation phase in this iteration as a representative use case for the user study. This includes actions such as finishing the meal, eating, and cleaning up.  While the meal preparation process varies significantly between households due to personal preferences and habits, post meal tasks tend to follow more standardized routines and provide assistance with step-by-step instruction. For example, these tasks often involve routine-based contexts (e.g., cleaning instructions) and situational contexts (e.g., turning off the stove). This focus enables the low-technology design of a context-aware system that supports common user needs while personalized to different household environments.
This phase is a relatable scenario to engage participants and gather targeted feedback on our prototype design. Importantly, this example was not intended to limit the prototype’s scope, but rather to provide a focused entry point for evaluating its functionality, which remains applicable across all phases of meal activities.

\begin{figure*}
  \includegraphics[width=\textwidth]{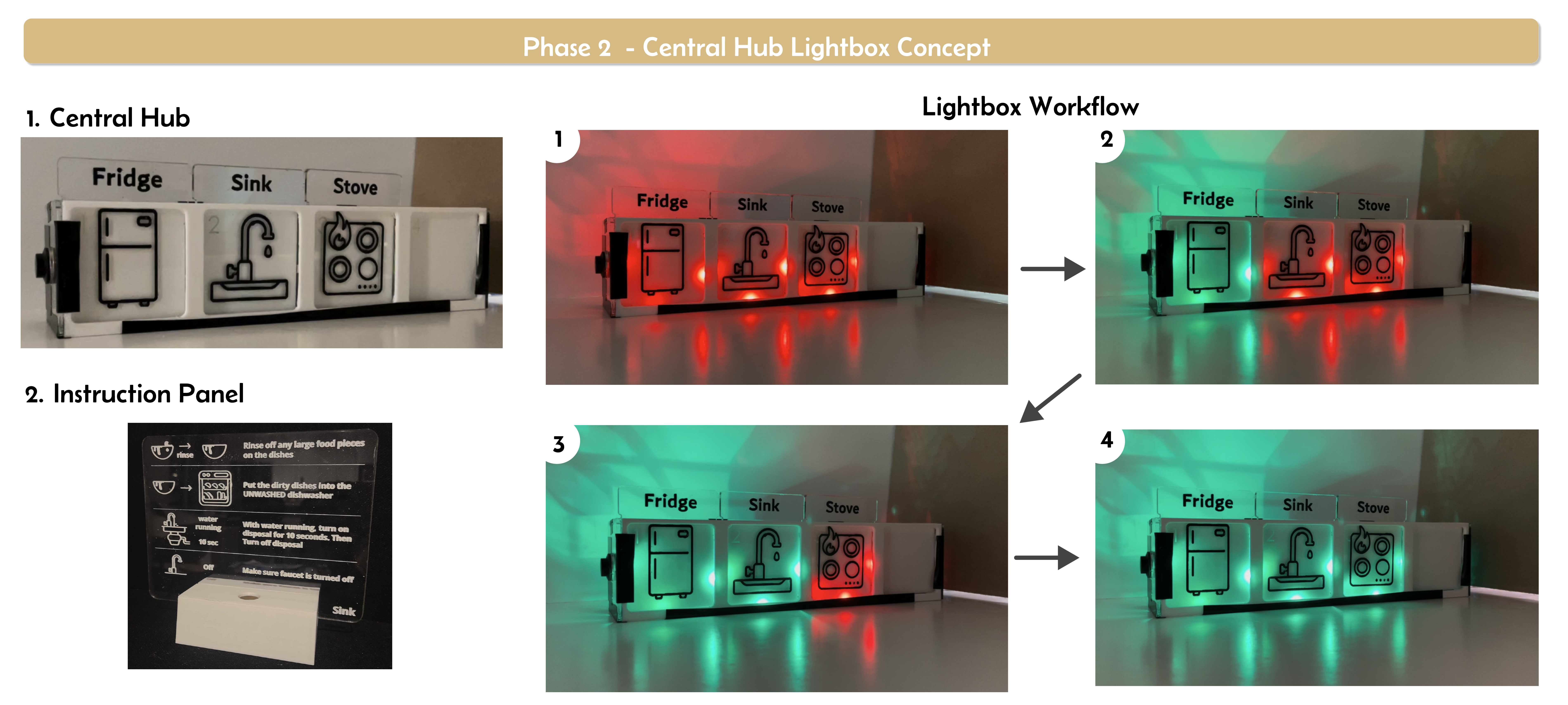}
  \caption{The Central Hub Lightbox includes two components: the central hub and the instruction panel. The central hub includes up to four customizable slots, enabling users to assign reminders to specific kitchen areas, with the flexibility to use only as many slots as they need. The central hub features icons representing different locations. On the right, the lightbox workflow is illustrated, where red indicates tasks in progress, and green indicates completed tasks.}
  \Description{The figure displays the Central Hub Lightbox with two main components: the central hub and the instruction panel. The central hub is shown on the left, with icons representing different locations (fridge, sink, and stove). Below the central hub, the instruction panel provides step-by-step guidance for tasks at specific locations. On the right, a lightbox workflow is depicted in four steps. The workflow uses colored lights: red indicates tasks in progress, and green indicates completed tasks. Step 1 shows all tasks as red. Step 2 highlights one task transitioning to green. Steps 3 and 4 show additional tasks turning green as they are completed.}
  \label{fig:phase2}
  \Description{The figure illustrates the Central Hub Lightbox Concept, which consists of two main components: the central hub and an instruction panel. The central hub, shown on the left, includes up to four customizable slots, enabling users to assign reminders to specific kitchen areas, with the flexibility to use only as many slots as they need. The central hub features icons representing three locations in the image: a fridge, a sink, and a stove. The instruction panel, displayed below, provides detailed steps for completing tasks related to each location, such as rinsing dishes or using the dishwasher. On the right, the lightbox workflow is depicted in four sequential images. In Step 1, all icons are illuminated in red, indicating that no tasks have been completed. In Step 2, the sink icon transitions to green while the fridge and stove remain red, signifying progress on the sink-related task. In Step 3, the fridge icon also turns green, showing that its associated task is completed. In Step 4, all icons are green, indicating that tasks for the fridge, sink, and stove are fully completed. The workflow uses colored lights: red indicates tasks in progress, and green indicates completed tasks. }
\end{figure*}

\subsection{Central Hub Lightbox Concept}
The device has two key components: a central hub box and instruction panels (shown in Figure~\ref{fig:phase2}). The central hub box featured four slots, each corresponding to a key area of the kitchen. It was intended for installation on a kitchen wall near the light switch. Participants could insert icons into these slots to represent the various tasks associated with each area, creating a clear visual representation of the meal preparation workflow. In the lightbox workflow shown in Figure~\ref{fig:phase2}, example areas include the fridge, stove, and sink faucet. The instruction panels were designed to be installed at the designated location in the kitchen. 
To begin using the system, users press a button on the central hub box to activate it. Once activated, the task icon on the hub illuminates in red, along with a corresponding instruction panel. For instance, if the fridge icon is the first to light up, the instruction panel near the fridge also lights up in red, guiding the user to complete their task at that station. After finishing the task, the user presses a button on the corresponding instruction panel to indicate completion, and the lights for both the icon and the instruction panel turn green. This action triggers the next task icon on the hub to illuminate in red, along with its associated instruction panel. The user would continue this process, moving sequentially through the tasks until all were completed. This design breaks down the tasks and provides step-by-step guidance.

\subsection{Participants and Group Feedback Procedures} 
In this study phase, we conducted a 45-minute in-person group feedback session at the common kitchen area of the same senior lifestyle program, CEP, as in Phase 1. The session consisted of about 10 pairs of older adults with MCI and their care partners. Due to the need to maximize the limited time given by the program, we did not collect detailed demographic information.   The session began with an overview of the study, introducing our research focus and demonstrating the new concept, which included a central hub and individual instruction panels. Participants were then given a brief tryout session with the system to provide initial feedback on its usability and functionality. Researchers observed and recorded data through observational notes and participant quotes to capture user interactions and feedback.
A participant was invited to do a simulated post-meal cleanup task with the device. Tasks included cleaning dirty dishes in the sink, clearing trash and food from the counter, ensuring the faucet was turned off, and checking that the refrigerator door was closed. Following the tryout, participants engaged in small group discussions, with 5–6 participants per group. Each pair or individual received a flyer listing sample tasks to facilitate discussion. A moderator in each group guided the conversation, posed targeted questions, and took notes on participant feedback on system functionality, usability, and potential improvements. Data collected during the session will be analyzed to identify key usability challenges and design opportunities.

\subsection{Results}
\subsubsection{Insight from the Central Hub Lightbox Concept}
During the session, participants interacted with the central hub design, which included a central hub box and instruction panels. We received feedback that participants appreciated the modular system’s ability to be fully customized to suit different scenarios and routines, aligning with \textbf{\textit{routine-based contexts}}. For example, participants suggested organizing the panels into breakfast, lunch, and dinner sections to better align with their meal preparation routines.

One of the participants proposed enhancing the central hub’s functionality by integrating a recipe database or detailed instructions for specific appliances. They explained, \textit{``If the goal is to allow a person with MCI to prepare meals independently, it could be actively associated with a recipe database so you can pick one recipe and follow the steps.’’} Furthermore, they recommended creating dedicated lists for different appliances. As they also noted, ``You could have a separate list for different appliances even the instructions will be the same for each appliance all the time. For example, there's a list for dishwasher with all instructions and things to keep in mind, and then you can switch to the toaster list for other instructions.’’

Participants also highlighted the need for specific \textbf{\textit{real-time contexts}} to enhance task management. While the participants appreciated the system’s ability to track different tasks, they pointed out the limitations of the instruction panel, which currently provides static step guidance. To address this, participants suggested adding features like a checklist or an automated system to track completed steps dynamically, which would help reduce cognitive load and improve task organization. Additionally, participants highlighted the need for real-time assistance in specific scenarios, such as reminding them to return items to their original locations. Participants proposed using sensors connected to objects to guide them in placing items back where they belong, addressing memory challenges common among older adults with MCI.

Safety reminders remain the top priority for participants, requiring both \textbf{\textit{real-time and situational contexts}} to effectively address their needs. Participants stressed the need for assistive systems to focus on urgent safety tasks, such as monitoring appliance statuses—like ensuring the stove is turned off—and responding promptly to potential hazards. To address these safety concerns, participants suggested integrating sensors with instruction panels and appliances. For instance, stove panels could be linked to smoke detectors, allowing the system to automatically display safety reminders or issue warnings when risks are identified. Similar to Phase 1, participants proposed multi-modal reminders, combining voice and visual alerts, to ensure critical messages are communicated effectively, even when they might be distracted or have sensory limitations. The voice reminders can also distinguish between urgent and non-urgent matters. Additionally, participants expressed that reminders should prioritize safety-related issues first, followed by reminders that support functional independence for older adults with MCI, and finally, features aimed at convenience for meal preparation.

Participants also provided valuable feedback on the central hub lightbox concept, focusing on its placement, interaction design, and content presentation. Firstly, participants emphasized the importance of clear and intuitive placement for the central hub and instruction panels. Researchers observed that the prototype’s interaction design was too complex, requiring multiple steps to interact, causing confusion about the order of interaction. For example, we observed one participant was uncertain about which individual instruction panel to interact with first and often skipped looking at the panels until prompted by their care partner or researchers.

Secondly, participants also discussed when the device should provide support. While the instruction panels displayed the necessary steps, participants found consolidating all tasks on a single board overwhelming and too wordy, as it required significant cognitive effort to read and remember the instructions. They recommended incorporating a checklist feature that allows users to check off completed tasks, which would make the interface more user-friendly and reduce cognitive load, particularly as they are already facing memory difficulties. Participants suggested reducing text, increasing visual contrast, and adopting a more intuitive color scheme to reduce the readability difficulties further. For example, dynamic icons that change color to indicate task progress were proposed as a potential improvement. Audio output reminders were also recommended to enhance accessibility, though participants noted the potential for irritation or habituation with repeated use.

Another area of concern was the content of the instruction panels. Participants emphasized the need for concise, accurate, and easy-to-understand instructions that cater to the needs of older adults with MCI. Some tasks on the panels were deemed too vague, making them difficult to comprehend. For example, one participant asked their care partner to explain the meaning of a fridge icon, demonstrating that even simple graphical illustrations might not be clear enough. Participants suggested including more explicit and detailed instructions to guide them through each task effectively.

Additionally, participants highlighted the need to simplify interactions. They suggested replacing individual buttons on each instruction panel with a single button to check in for all tasks, reducing the cognitive effort required to operate the system. Automatic activation and deactivation of the device were also preferred, as participants noted that remembering to turn devices on or off could be challenging for older adults with MCI.

\begin{figure*}[t!]
  \includegraphics[width=\textwidth]{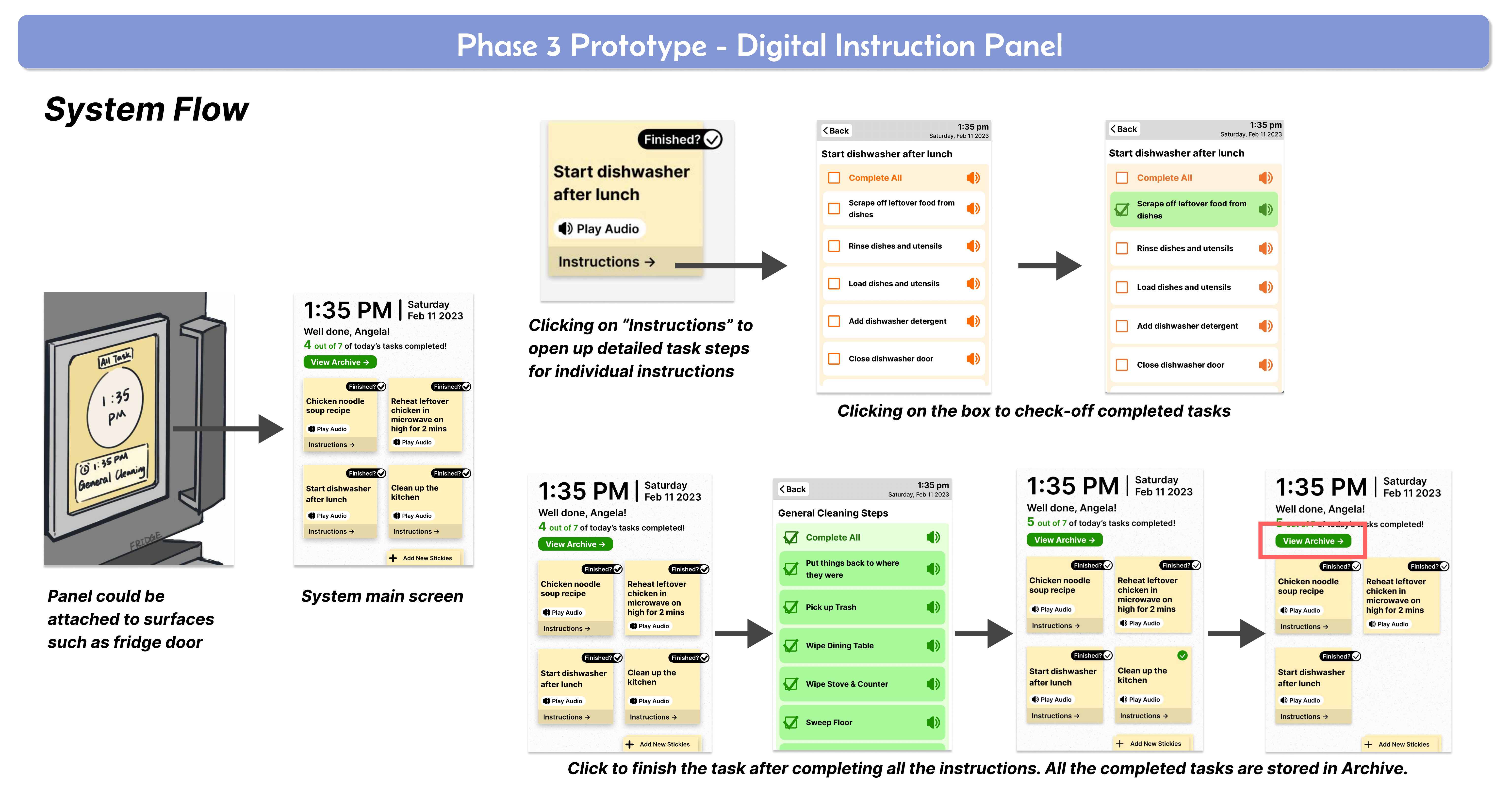}
  \caption{The system flow of the Digital Instruction Panel. The panel, designed to be attached to surfaces like a refrigerator door, features a main screen displaying tasks for the day. Users can access step-by-step instructions for each task by selecting ``Instructions.'' Completed tasks can be checked off, and once all tasks are finished, reminders are archived for future reference. Each note or instruction also includes the option to play audio to have it read aloud.}
  \Description{The figure illustrates the system flow of the Digital Instruction Panel. On the left, a panel is shown attached to a refrigerator door. The main screen of the panel displays a list of tasks with associated buttons to play audio or access instructions. A step-by-step process is shown. The users can tap ``Instructions'' to view detailed task guidance. Users can check off completed tasks using checkboxes next to each step. Once all tasks are completed, users can tap ``Finished,'' and the system moves the completed tasks into the archive, accessible through a ``View Archive'' button on the main screen.}
  \label{fig:phase3}
\end{figure*}

\section{Phase 3: Refining and Testing with a Technology Probe}
\subsection{Design Process}
Our second phase highlighted the need for a system that is ``always-on'' and personalized for the users such that the information was meaningful to them and easily followed. While the placement of instruction panels near each task station was also implemented to enhance convenience, we observed that some participants were confused by the interaction flow. Users also talked about the need for multi-modal reminders and dynamically modify the reminders based on the status of the house. All these requests are not achievable by simple low-tech approaches. 

We revisited notes from both Phase 1 and Phase 2 and found that some participants were already familiar with different technologies and constantly compared our system to smart speakers and mobile phones. To ensure we can address the diverse participants' need, we embraced a digital interactive solution. We also envision a system that is tied into a smart environment and can provide situational-based context -- reacting to the fridge doors being left open and detecting the start and end of meal preparation activity.

\subsection{Digital Instruction Panel}
We envisioned a Digital Instruction Panel that served as a digital hub for reminders and task management. The design of the Digital Instruction Panel is shown in Figure~\ref{fig:phase3}. 
The main screen displayed the date, time, and digital sticky notes, each containing a brief reminder description. For accessibility, each sticky note included a ``play audio'' button that read the reminder aloud for older adults with MCI who preferred audio assistance. A ``mark as done'' button also allowed them to indicate when a task or reminder had been completed. For reminders that required detailed instructions, the sticky notes included a clickable bottom section. When selected, this section expanded to show a detailed list of task steps linked to the reminder. Each step included a checkbox, allowing users to mark tasks as completed as they progressed through the instructions.
This design leveraged the interactive capabilities of the Digital Instruction Panel to provide a user-friendly, flexible, personalized solution. We also envisioned the system to be tied into a smart home sensor system to enable situational-context support, by providing reminders that react to the state of the house. The system aims to offer tailored user preferences and requirements through visual, tactile, and auditory features, ensuring an accessible and engaging experience for older adults with MCI. 

\begin{figure}
  \includegraphics[width=\columnwidth]{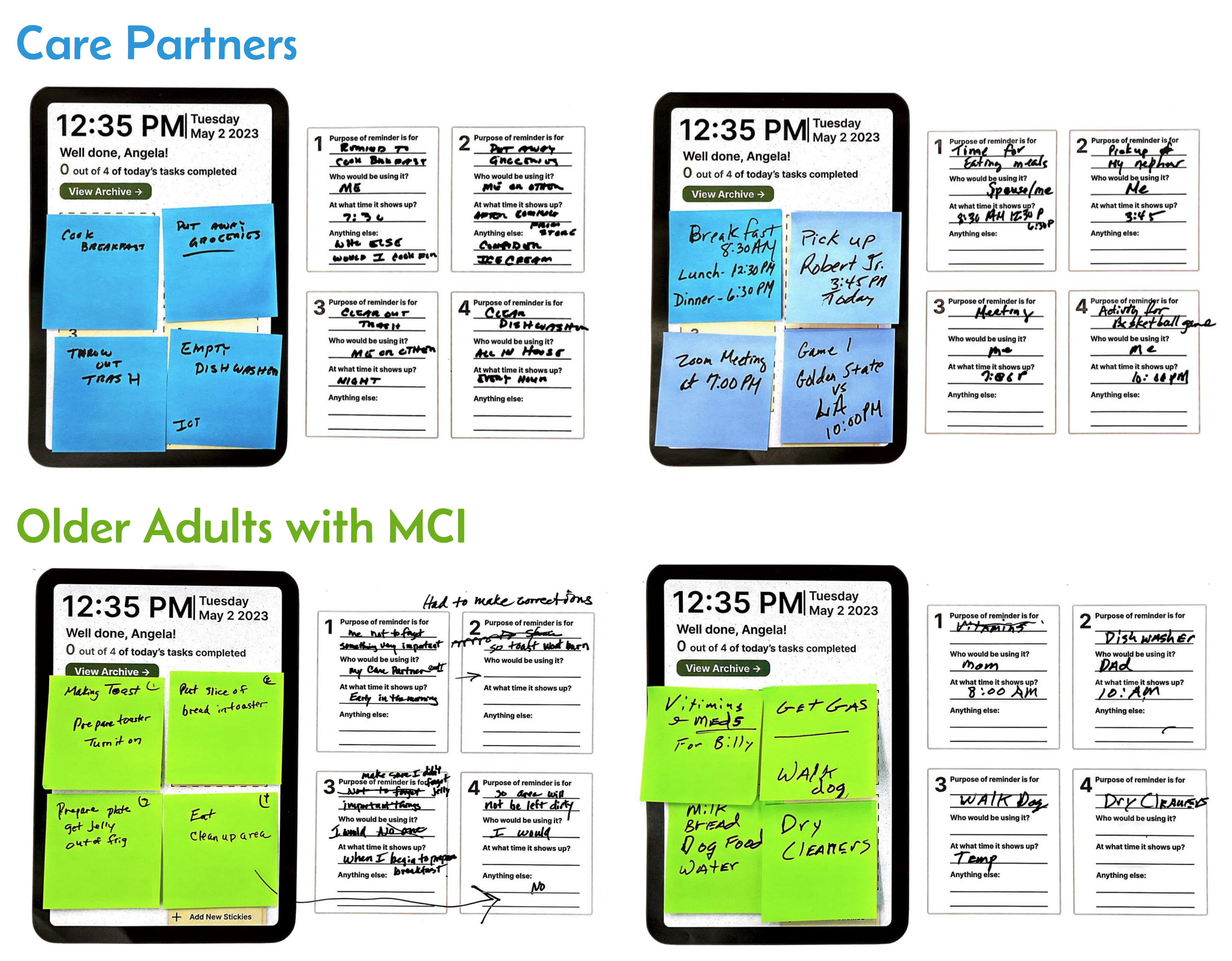}
  \caption{Both care partners and older adults with MCI completed the worksheet during the brainstorming session. Responses from care partners are shown in blue, while those from older adults with MCI are shown in green. Participants were asked to specify reminders they would display on the screen. Alongside, they answered questions regarding the purpose of the reminder, who would use it, when it should appear, and any additional notes they had.}
  \Description{The figure displays worksheets completed by two groups: care partners (blue sticky notes) and older adults with MCI (green sticky notes). Each worksheet contains a Digital Instruction Panel mockup at the left with tasks entered on sticky notes and a set of written responses on the right. The sticky notes represent reminders participants suggested for display on the Digital Instruction Panel. The written responses address questions about the purpose of the reminder, the intended user, display timing, and additional comments.}
  \label{fig:phase3_participants}
  \Description{The figure illustrates worksheets completed by two groups of participants during a brainstorming session: care partners and older adults with MCI. Responses from care partners are represented in blue, while responses from older adults with MCI are shown in green. Each worksheet consists of a digital mockup screen at the top, where participants added sticky notes specifying reminders. Care partners included tasks such as ``Cook breakfast,'' ``Put away groceries,'' ``Throw out trash,'' and ``Empty dishwasher,'' while older adults with MCI added reminders like ``Make toast,'' ``Put slice of bread in toaster,'' ``Get gas,'' ``Walk dog,'' and ``Dry cleaners.'' Next to the digital mockup screen, participants answered structured questions about the purpose of each reminder, the intended user, the timing of when the reminder should appear, and any additional notes or context.}
\end{figure}

\subsection{Participants and Group Feedback Process}
To iterate on our design and gain feedback on how participants will use the system, we conducted another group feedback session through the senior lifestyle program, CEP.
% We showed them Figma prototype of the device
We conducted one 45-minute moderated session with 19 participants, including 11 older adults with MCI and nine care partners. During the session, participants were introduced to our project and the Figma prototype of the system. A small group activity was conducted, where participants brainstormed use cases for the system and provided feedback on the prototype. Each small group will be guided by an assigned researcher. During the discussion, the researcher observes and notes the feedback accordingly.
Figure~\ref{fig:phase3_participants} shows the worksheet completed by the participants during the brainstorming session.

\subsection{Group Feedback Results}
From the small group activity, most older adults with MCI and their care partners focused on reminders related to routine tasks, safety reminders, and appointment reminders. These reminders were often specific about when they should appear, reflecting the participants’ desire for timely and actionable prompts. Care partners’ worksheets also revealed an interesting observation that they added reminders for their own rather than exclusively setting reminders for the older adults with MCI. This observation aligns with the prototype's dual end-user design goal, which aims to serve both older adults with MCI and their care partners effectively.

Discussions on the prototype among participants centered around two themes. The first theme was the functionalities and features of the Digital Instruction Panel. Participants appreciated the familiar and intuitive sticky note format but questioned how new reminders or information would be added to the system. Suggestions included features like auto-detection of completed tasks to reduce the cognitive and physical effort required to manage the device. They also proposed additional interaction methods, such as speech-to-text capabilities, digital writing tools like an Apple Pencil, and scanning physical sticky notes directly into the system for greater flexibility. Furthermore, participants emphasized the value of linking the system with their phone calendars, enabling reminders, such as grocery lists, to be accessible across multiple platforms for convenience. Participants expressed a desire for the system to extend its functionality beyond a single location, suggesting it provides reminders in different rooms throughout the home or even outside the home. They also raised questions about how notifications could effectively function in different locations.

The second theme was around the usability and learning curve of the Digital Instruction Panel. Participants noted that individuals already familiar with technology might find the system helpful, but those less experienced with tech perceived it as ``another gadget'' that required effort to learn. This raised concerns about whether the learning curve might outweigh the tool’s functionality, especially for older adults with MCI. Participants also highlighted the potential for the device to feel overwhelming, particularly when users are required to input a large amount of information at the initial setup. Discussions also touched on the cognitive changes experienced by people with MCI, emphasizing the need for the system to accommodate these changes to suit their specific needs. Suggestions included making sticky notes adaptable, with flexible levels of detail that can be adjusted as memory declines. 

Given the feedback from participants, we were confident that we have identified a design that is potentially useful for participants. We implemented a working version of the digital interface on an iPad and aimed to evaluate it in a more ecologically valid setting. 

\subsection{Participants and Prototype Walkthrough}

\begin{figure}[t]
    \centering
    \setlength{\belowcaptionskip}{-10pt}
    {
        \label{fig:kitchen}
        \includegraphics[width=\columnwidth]{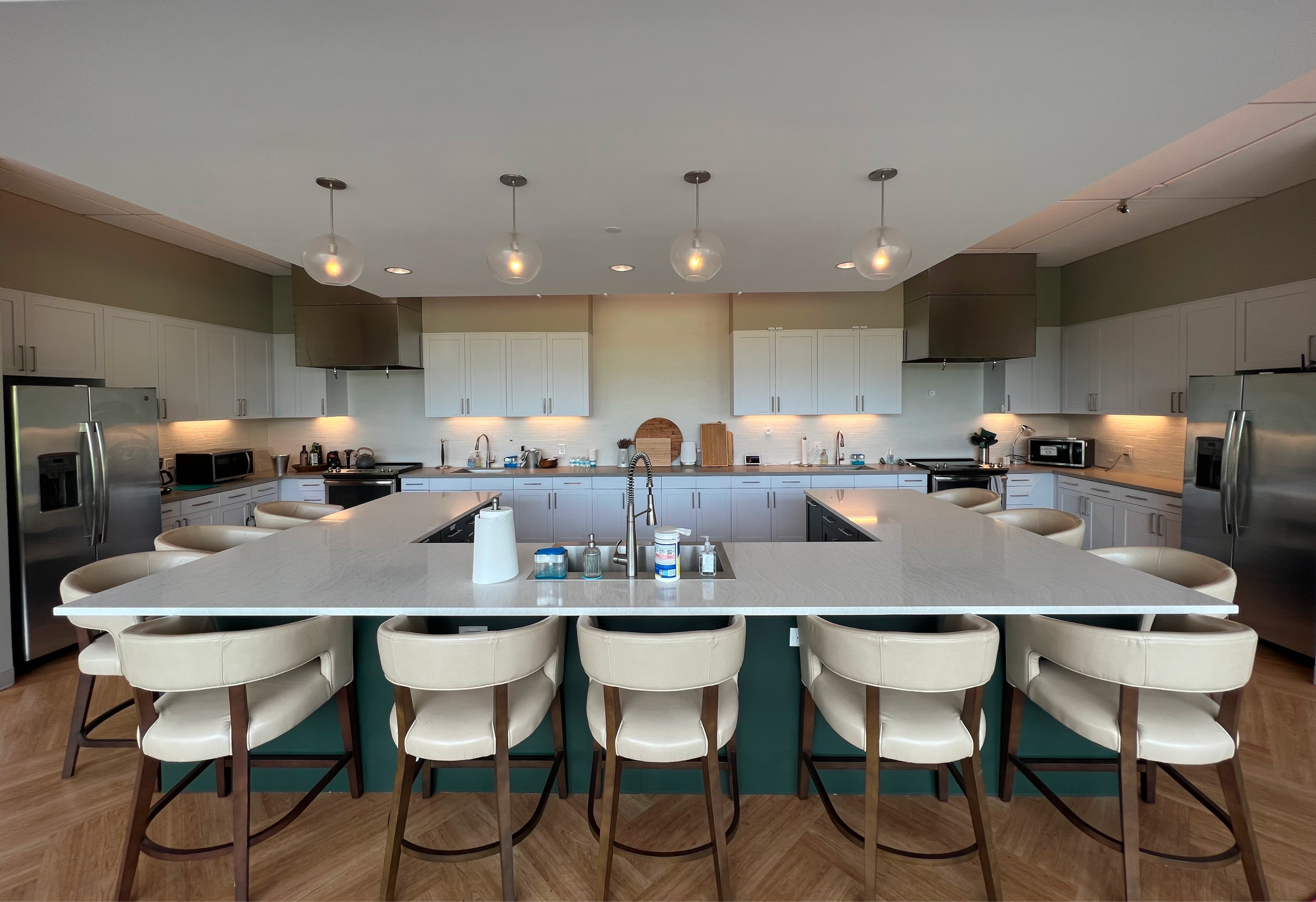}
    }
    \hfill
    {
        \label{fig:fridge}
        \includegraphics[width=\columnwidth]{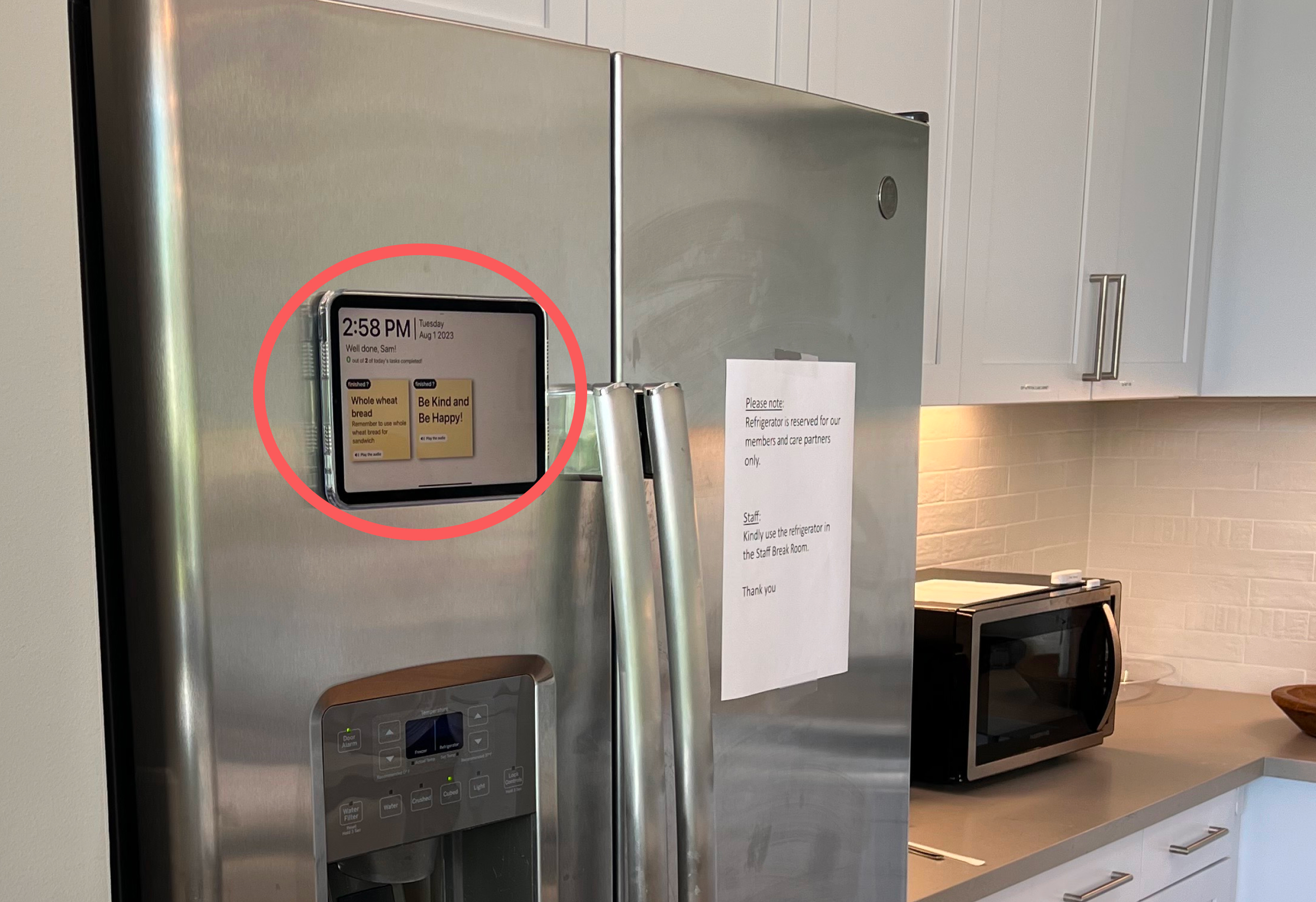}
    }
    \caption{Our study was conducted in the common kitchen area of the senior lifestyle program. The top image shows the kitchen setup, while the bottom image shows the fridge setup with the Digital Instruction Panel attached to the refrigerator (circled in red).}
    \label{fig:setup}
    \Description{Our study was conducted in the common kitchen area of the senior lifestyle program. The top image shows the kitchen setup, while the bottom image shows the fridge setup with the Digital Instruction Panel attached to the refrigerator (circled in red).}
\end{figure}

In this phase of the study, we conducted prototype walkthroughs and semi-structured interviews with six participants recruited thro\-ugh the senior lifestyle program. The study was conducted in the common kitchen area of the senior lifestyle program (shown in Figure~\ref{fig:setup}). All participants consented to the session, which was audio-visually recorded for analysis. We collaborated with other researchers to have our walkthrough be a ``distractor'' task between the two conditions of a controlled user experiment. Participants were introduced to the background and motivation behind the application prototype. The researchers then provided a tutorial on the system’s functionalities, explaining that care partners could send reminders and how digital sticky notes are displayed. Participants were also shown how to interact with the sticky notes, including viewing task instructions and marking tasks as completed. Following the tutorial, participants were tasked with preparing a peanut butter and jelly sandwich, a familiar and simple meal preparation task. During the task, the researcher used a Wizard of Oz setup to simulate the system’s situational-context functionality, sending digital sticky notes in real-time, such as reminders to close the microwave, use wheat bread when they first enter the kitchen and clean up the workspace after they completed making the sandwich. The full interaction with the system prototype was around 2 and 10 minutes. After completing the task, participants took part in a post-task interview to provide feedback on the application’s overall user experience.

\subsection{Prototype Walkthrough Results}
\subsubsection{Insight from the Digital Instruction Panel}

Participants saw the potential of the Digital Instruction Panel to support \textbf{\textit{routine-based contexts}} by integrating recurring habits and daily routines. For example, P3-3 noted that the system could be used for medication reminders and appointments, reducing the need to rely on calendars. Similarly, P3-4 expressed a desire for the reminder to integrate personal calendar events and other important tasks, enabling them to access everything in one place. Routine-based contexts for older adults aren’t solely about practical tasks where they can also include routines that promote emotional well-being. We observed that P3-6 responded with a smile to motivational reminders like ``Be kind and be happy,'' which shows the potential of the emotional value of such messages that could add to routine support.

While working on the meal preparation task, we observed inconsistency in interaction with \textbf{\textit{real-time}} notifications and reminders. P3-6 showed actively checking reminders before performing steps to ensure accuracy, while P3-3 used the reminders as post-meal preparation to verify completed tasks. In contrast, P3-1 and P3-5 noticed the initial reminders but forgot to check on the rest reminders, and P3-2 and P3-4 missed all reminders entirely. This observation shows the need for improvements in how the Digital Instruction Panel processes and delivers real-time contextual reminders, ensuring timely and consistent delivery of notifications that capture users’ attention during task execution. 

Participants shared how they envisioned themselves using \textbf{\textit{situational contexts}} to enhance their daily routines. For example, P3-3 saw themselves benefiting from the Digital Instruction Panel as a reminder to close the fridge door, while P3-5 imagined using it to help prepare something later based on a partner’s instructions.

Participants also provided additional feedback on the design of the Digital Instruction Panel, offering suggestions for improvement. First, P3-5 mentioned that the Digital Instruction Panel's placement on the fridge did not align with their workflow during tasks and was not easily visible. In contrast, P3-3 was excited about the potential to place the Digital Instruction Panel in other locations, such as their bathroom or upstairs, showing the importance of tailoring the device’s placement to fit personal routines.

In addition to placement concerns, we also observed participants experienced usability challenges with the interface. For instance, P3-6 accidentally pressed the wrong button and struggled to clear completed tasks, showing the need for a more accessible design to reduce errors and improve user experience. P3-6 also relied heavily on the reminders but found the lack of clear instructions, such as a sign or notification, to mark task completion and ensure the system supports users throughout the entire task flow.

Participants also began considering ways the prototype could be used remotely. For instance, P3-1 expressed interest in understanding how to activate the device and send sticky notes, particularly whether these could be sent remotely when the user was away from home. This shows their excitement about extending the prototype’s functionality to support remote interactions, aligning with our design intent of enabling care partners or users to engage with the device from different locations.

\section{Discussion and Implications for Design}

\subsection{Remaining Gaps and Challenges}
Overall, the types of contexts we identified across all three design iterations--routine-based, real-time, and situational--serve as a basic background for creating a context-aware, personalized assistive device tailored to older adults with MCI that aligns with their cooking habits, preferences, and environments.
However, challenges remain in fully addressing the diverse contexts that need to be recognized for older adults with MCI. For example, emotional contexts, such as stress or frustration during meal preparation, could impact how users interact with the system and their willingness to rely on it. 
Similarly, our situational context is only limited to appliances and objects in the kitchen and ignores the social situational context.
The presence of care partners or family members could influence the way reminders and guidance are delivered. 

In our walkthrough, participants were asked to make a peanut butter and jelly sandwich -- a scenario without any urgency or safety concerns. Throughout our design process, participants consistently raise the desire for reminders during unsafe scenarios such as open fire. How the system should deliver urgent reminders remains an open question. This is especially important as some participants missed reminders entirely due to the reminder tone being too soft or insufficiently noticeable. The alert system should ensure that reminders are attention-grabbing and align with its urgency. For instance, reminders related to health or safety, such as turning off the stove, should use more prominent alerts, such as louder tones or visual signals. Less urgent reminders, like completing a meal preparation step, could use subtler methods to avoid overwhelming or irritating the user.

\subsection{Reflection on the Design Process and Implications for Future Design}

\subsubsection{Balancing Simplicity and Technological Integration for Older Adults with MCI}
Our initial design approach emphasized simplicity, aiming for a low-tech solution to address the unique challenges faced by older adults with MCI. Prior research shows that older adults without cognitive impairments often require more time to adapt to new technologies compared to younger generations~\cite{pradhan2020use, vines2015age}. For those with MCI, this learning curve becomes even steeper, posing additional challenges in everyday tasks such as using ticket machines or managing online banking~\cite{rosenberg2009perceived, malinowsky2010ability}. With this in mind, we designed a lightbox prototype with button-based functionality, intentionally keeping the interface minimalistic. The goal was to create an accessible and easy-to-use tool that seamlessly integrates into users' meal preparation routines without requiring significant changes to their habits or introducing a demanding learning process.

Through our interviews, we found simple low-tech solutions were not able to meet the dynamic and varied needs of older adults with MCI and their care partners. During interviews about their meal preparation habits, some participants mentioned their existing habit of using technologies like Google Home or Amazon Alexa. These technologies have become a part of their routines, serving as tools such as voice-activated reminders and customizable notifications. Participants often compared our prototype to these tools, emphasizing gaps in functionality and suggesting features they wished to see integrated into the system. For instance, while participants appreciated the simplicity of the lightbox, they expressed a desire for voice commands, real-time notifications, and greater adaptability to different tasks and contexts.

Reflecting on our design process, we learned that designing a context-aware meal preparation assistance for older adults with MCI requires not only meeting their needs but also integrating seamlessly into their existing technological context. This does not mean reverting to familiar technologies or metaphors, but engaging the users to understand their current technology routines and familiarity. Therefore, it is important for future design to balance simplicity with technological adaptability for creating effective assistive systems. While our initial low-tech approach provided a solid foundation, overly simplified designs may not fully address the dynamic needs of older adults. New system should complement or add value to the familiar tools users already rely on.

\subsubsection{The Role of Context Aware}
Our phase 2 central hub lightbox system relies on manual activation and static reminders.
Participants noted that if users remembered to press the button, they often recalled the task itself, raising questions about the system's necessity. Conversely, we can also imagine some users might activate the system but forget its purpose, leaving tasks incomplete or creating potential safety risks. This made us embrace situational context and had our phase 3 device to be reactive to the state of the house. For example, having a reminder for closing the fridge door if the door is left open. Understanding the context where reminders are needed will better align the system with the user's real-world needs. 

Throughout the design process, safety-related needs remain the most frequently mentioned across all three phases, including kitchen safety, cooking safety, and personal health safety, aligning with findings from ~\citet{johansson2011cognitive}. In the Phase 1 interview, we learned that some older adults with MCI were not allowed to use stoves by their caregivers due to safety concerns. While they shifted to using microwaves to avoid open flames, safety risks still persisted because they often forgot the instructions. As they struggle to perform daily tasks independently and require more external support, it reduces their confidence and limits their ability to maintain social connections and control over their routines, undermining the benefits of aging in place. Participants expressed a desire for assistive devices that offer real-time detections and reminders to ensure safety during meal preparation. By recognizing routine-based, real-time, and situational contexts, context-aware assistive technology could empower older adults with MCI to perform meal preparation tasks independently once again.

The context influences not only which reminders are needed, but also how they are delivered. In Phase 3, we observed the challenges that older adults with MCI face in recognizing or hearing notifications, revealing the need for proactive monitoring and real-time adaptation to ensure successful delivery of the reminders, even when users do not fully engage with the system. At the same time, the system needs to be aware of its past reminders and balance its effectiveness with user comfort, as repeated notifications might lose their impact over time, while intrusive alarms or sounds could irritate household members. 

\subsubsection{The Role of Care Partners}
Care partners also play a key role in shaping the design process by identifying essential contexts that the system should recognize. As primary caregivers, they spend significant time assisting older adults with MCI. For example, in the Phase 1 habit interview, P1-3 shared that when the older adult forgot to close cabinet doors, the care partner would ``just come along behind and close it, ''  highlighting one of the many tasks care partners routinely manage. They were also the ones who were inundated with reminding the older adults. A successful reminder system is as much a tool to assist older adults with MCI to live independently as is to lighten the burden of the care partner.  

By involving care partners in the design, we gained valuable insights into the types of information and reminders the system should provide, customized for the older adult with MCI. This approach ensures that the system addresses immediate needs highlighted by the care partners while supporting the independence of older adults. Incorporating care partner input into the design process not only reduces their workload but also fosters a balanced caregiving dynamic, enabling both older adults and their care partners to benefit from the assistive system.

\subsubsection{Power Dynamics and Participant Agency in Design Process}
Our study involves older adults with MCI, who may sometimes face challenges articulating their needs and providing feedback during the design process.
This might lead to designers bringing preexisting notions of the participant’s wants.
Balancing this power dynamic between designers and participants is often a common challenge when working with marginalized populations~\cite{goodwill2021beyond, jiang2022understanding, yoo2024remembering}.
We employed a few strategies to ensure that their opinions were heard and valued. First, we included care partners in the iterative design process to help gain perspectives that participants might not have noticed on their own. We also grounded our first prototype in prior research, using low-tech, tangible interactions (e.g., physical buttons and lightboxes) to lower cognitive barriers for older adults with MCI. Despite these efforts, some participant feedback was not fully integrated into the design at certain stages. For example, although voice-based reminders were suggested early in the process, we initially prioritized visual and low-tech solutions in Phase 2. This decision was influenced by the challenges of implementing voice output in a low-tech setup. We implemented a voice instruction feature in Phase 3 after repeated feedback. We acknowledge our approach was imperfect and likely empowered or constrained some participants. Our experience serves as an example of why designers and researchers should carefully consider how power is distributed in the iterative design process, how participant input is interpreted, and how to support users' agency.

\subsection{Limitations and Future Work}
One key limitation of our study was the recruitment scope, as participants were drawn from the senior lifestyle program \anon{CEP}. Participants are those who are capable of joining the program and do not have other work/personal commitments. This limited the diversity of the participant pool and may have reduced the generalizability of our findings.
Furthermore, due to the study's extended timeline, we were unable to retain the same group of participants across all three phases. This approach could have introduced variability in feedback and insights across iterations, potentially limiting our ability to fully track participant perspective changes or assess our prototypes' longitudinal effectiveness. 
Future studies should expand the participant population to include a more diverse demographic, offering broader insights into the needs and preferences of older adults with MCI and their care partners. Although each phase involved a limited number of participants, our findings provide a foundation for researchers developing context-aware assistive technologies for meal preparation.

In developing the digital screen prototype, we used a Wizard of Oz approach, assuming that relevant contexts, such as activity recognition and object states could be effectively captured by sensors in the future. Future work should address the challenges of designing robust context-aware systems that can reason about the reliability and accuracy of sensor data. Inaccuracies in these systems could compromise their ability to provide timely and relevant assistance. Privacy concerns associated with collecting and processing user data should also be considered, particularly in home environments. 

\section{Conclusion}
In this study, we address the challenges of understanding the contexts and the design considerations that are essential for assistive technologies for older adults with MCI in meal preparation. Through an iterative design process that incorporates insights from both older adults with MCI and their care partners, we progressively refined our prototype, evolving from a low-tech lightbox to a context-aware personalized system tailored to their specific needs.

Based on the gathered feedback, we have identified three critical contexts--routine-based, real-time, and situational--that are essential for developing adaptive meal preparation assistive systems for this population. Reflecting on our design process, we learned that effective assistive technologies must involve both older adults with MCI and their care partners, as care partners provide invaluable insights into the contexts and needs of older adults. Assistive technologies should not only adapt to the evolving needs of older adults but also support care partners by reducing their workload and fostering collaborative caregiving. Additionally, while older adults face cognitive decline, low-tech solutions are not always the most effective approach, and the new technology should build upon their existing technology routine.

\begin{acks} 
We thank the participants and the Cognitive Empowerment Program for making this project possible. This work is funded by the National Science Foundation Grant (IIS-2112633)
\end{acks}

% \section{Appendices}

% If your work needs an appendix, add it before the
% ``\verb|\end{document}|'' command at the conclusion of your source
% document.

% Start the appendix with the ``\verb|appendix|'' command:
% \begin{verbatim}
%   \appendix
% \end{verbatim}
% and note that in the appendix, sections are lettered, not
% numbered. This document has two appendices, demonstrating the section
% and subsection identification method.

\bibliographystyle{ACM-Reference-Format}
\bibliography{main_reference}

\end{document}